\documentstyle[prl,aps,psfig,epsf,twocolumn]{revtex}
\begin{document}
\draft
\twocolumn[\hsize\textwidth\columnwidth\hsize\csname@twocolumnfalse%
\endcsname

\preprint{}

\title{Bilayer Quantum Hall Systems: Spin-Pseudospin
Symmetry Breaking and Quantum Phase Transitions}
\author{Sankar Das Sarma$^1$ and Eugene Demler$^2$
}
\address{$^1$Department of Physics, University of Maryland,
College Park, MD 20742-4111\\
$^2$Physics Department, Harvard University,
Cambridge, MA 02138
}

\date{\today}
\maketitle
\vskip 0.5 cm
\begin{abstract}
We discuss and review recent advances in our understaning of quantum
Hall systems where additional quantum numbers associated with spin
and/or layer (pseudospin) indices play crucial roles in creating
exotic quantum phases. Among the novel quantum phases we discuss are
the recently discovered canted antiferromagnetic phase, the
spontaneous interlayer coherent phase, and various spin Bose glass
phases. We describe the theoretical models used in studying these
novel phases and the various experimental techniques being used to
search for these phases. Both zero temperature quantum phase
transitions and finite temperature phase transitions are
discussed. Emphasis in this article is on the recent developments in
novel quantum phases and quantum phase transitions in bilayer quantum
Hall systems where nontrivial magnetic ground states associated with
spontaneous spin symmetry breaking play central role.
\end{abstract}

\vspace{0.5 cm}

\vskip -0.2 truein
\vspace{1 cm}
]
\narrowtext

\section{Introduction}
In a seminal paper more than fifteen years ago Halperin pointed out
\cite{citation1} that if additional (i.e. in addition to the 2D
orbital motion) degrees of freedom or quantum numbers (e.g. spin,
layer, etc.) are important in quantized Hall systems, then
intriguing novel phenomena associated with new quantum phases
described by multicomponent (generalized Laughlin) wavefunctions, the
so-called Halperin wavefunctions, become theoretically possible. The
observation \cite{citation2} and the theoretical understanding
\cite{citation3} of the $\nu=1/2$ quantum Hall effect ( where $\nu$
indicates the {\it total} Landau level filling throughout this paper,
rather than the filling factor for each individual layer) 
was one of the spectacular
early confirmations of Halperin's ideas. During the last decade there
has been a great deal of theoretical and experimental work
\cite{citation4} on multicomponent quantum Hall systems with the
primary emphasis on $\nu=1/2$ and $\nu=1$ bilayer structures (with the
layer index serving as a pseudospin variable with $U(1)$ symmetry in
the absence of interlayer tunneling) although there have also been
some interesting developments in single layer composite spin states in
higher Landau levels \cite{citation4}.

Since excellent and extensive reviews of bilayer (spin-polarized) 
$\nu=1/2$ and $\nu=1$ quantum Hall physics already exist \cite{citation4}
in the literature, we focus in this article on the recently discovered 
\cite{citation5,citation6,citation7,citation8,citation9,citation10,citation11,citation12,citation13,citation14,citation15} rich and interesting physics associated 
with the bilayer $\nu=2$ quantum Hall systems, where both spin and
pseudospin dynamics compete in the quantum Hall phenomena leading to a
number of novel magnetic quantum phases (and interesting possible
$T=0$ quantum phase transitions among them as physical 
parameters such as interlayer tunneling, Zeeman
coupling, interlayer separation, etc. are tuned), some of which seem
to have already been experimentally observed
\cite{citation16,citation17,citation18,citation19,citation20}.

To introduce the concept of a spin symmetry breaking quantum phase
transition in bilayer quantum Hall systems we start from the schematic
non-interacting single particle energy level diagram shown in Fig.
1. Consider a bilayer system in an external magnetic field with a
total filling factor $\nu=2$ (i.e. the average filling in each layer
is 1 in the balanced situation under consideration) assuming there is
an interlayer tunneling induced symmetric-antisymmetric gap
$\Delta_{SAS}$ between the orbital levels and a magnetic field induced
Zeeman gap $\Delta_Z$ between the spin up and down levels. We restrict
ourselves entirely to the lowest Landau orbital level for the time
being, neglecting coupling to all higher orbital Landau levels
assuming $\hbar \omega_c$ to be very large. Therefore tunneling- and
spin-split energy levels are:
\begin{eqnarray}
E_{s\alpha} = s \frac{\Delta_Z}{2}+ \alpha\frac{\Delta_{SAS}}{2} 
\label{eq1} 
\end{eqnarray}
where we have ignored the ground state Landau level energy 
$\hbar \omega_c/2$.  The
spin ($s$) and the layer-pseudospin ($\alpha$) quantum indices are
each discreet and can take only two values $s=\pm1$ (up/down)
$\alpha=\mp 1$ (symmetric/antisymmetric). From now on we use notation
$1$ and $2$ to denote symmetric ($-1$) and  and antisymmetric ($1$)
states.
Equation (\ref{eq1})
therefore desribes four possible single particle energy levels (as
shown in Fig. 1) with the following energy values $E_{-1}=
-{\Delta_Z}/{2} -{\Delta_{SAS}}/{2}$; $E_{-2}= -{\Delta_Z}/{2}
+{\Delta_{SAS}}/{2}$; $E_{+1}= +{\Delta_Z}/{2} -{\Delta_{SAS}}/{2}$;
$E_{+2}= +{\Delta_Z}/{2} +{\Delta_{SAS}}/{2}$ where we use the
notation 1(2) and $-(+)$ to denote the symmetric (antisymmetric)
tunneling-split and spin up (down) levels respectively.
Each of these four single particle levels has a
macroscopic Landau degeneracy associated with the ususal magnetic
field induced Landau quantization which is not explcitly shown above. We
emphasize that all four levels belong to the lowest orbital Landau
level, and higher Landau levels are ignored in our consideration.
\begin{figure*}[h]
\centerline{\epsfxsize=8cm
\epsfbox{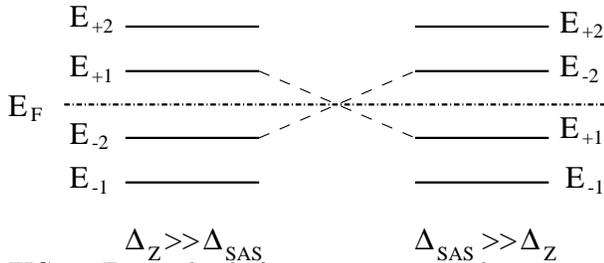}
}
\caption{Energy levels for non-interacting electrons
in bilayer quantum Hall system in the lowest Landau level
($\nu=2$) with $E_F$ as a Fermi level}
\label{f1.eps}
\end{figure*}

For total $\nu=2$ only two of these energy levels are occupied (
and the other two empty), as shown in Fig. 1 in the single particle
picture. Depending on whether $\Delta_Z > \Delta_{SAS}$ (or $
\Delta_{SAS} > \Delta_Z$) the weak (strong) tunneling situation, there
is always a single particle excitation gap
$\Delta_g=|\Delta_Z-\Delta_{SAS}|$ at the Fermi level. The $\nu=2$
bilayer system is thus always in an incompressible quantized Hall state
by virtue of the existence of the single particle excitation gap
$\Delta_g$. This is in sharp contrast to the corresponding extensively
studied \cite{citation4,citation21} zero tunneling $\nu=1$ bilayer
system in the fully spin polarized situation where the single particle
gap is by definition zero ($\Delta_{SAS}=0$ in the absence of
tunneling), and any incompressibility must necessarily arise from
interlayer many-body correlation in the pseudospin space. We thus
already note two significant differences between $\nu=2$ bilayer and (
the better known and well-studied) $\nu=1$ bilayer physics. In the
$\nu=2$ case, the system is always incompressible by virtue of the
existence of the non-zero excitation gap $\Delta_g \neq 0$ whereas in
the $\nu=1$ zero-tunneling case there is never any
single particle tunneling gap, but the $\nu=1$ system may be driven
incompressible (for suitable values of interlayer separation, etc.)
by interlayer coherence effects as in Halperin $(1,1,1)$ state
\cite{citation1}-\cite{citation4} or in related many-body incompressible
states \cite{citation21}; the other crucial difference between the two
is that the interesting coherence and quantum phase transition physics
in the $\nu=1$ case \cite{citation4,citation21} arise entirely from
the pseudospin correlation effects in the absence of tunneling since
the real electron spin is assumed to be completely frozen out by the
large Zeeman splitting in the system ( i.e. the electrons are are all
completely spin-polarized) whereas the interesting quantum phase
transition and symmetry breaking physics
\cite{citation5,citation6,citation7,citation8} in the $\nu=2$ bilayer case arise
from the interplay between the spin and pseudospin correlations (with
both $\Delta_Z$ and $\Delta_{SAS}$ playing important roles, and the
interesting regime is in fact $\Delta_{SAS}/\Delta_Z \sim 1$)
and the actual symmetry breaking in the $\nu=2$ bilayer case 
\cite{citation5,citation6,citation7,citation8,citation9,citation10,citation11,citation12,citation13,citation14,citation15} is a peculiar breaking of the
real spin symmetry (see Fig. 2) in the $x$-$y$ plane of the layers (with the
magnetic field orientation being along the $z$ direction in the usual
notation).

Of the four $E_{s\alpha}$ single particle states in the lowest Landau
level (Fig. 1), arising from spin and pseudospin splitting (Zeeman and
tunneling effects respectively), only two are filled with electrons
for total $\nu=2$.  The level $E_{-1}$ is always the lowest energy
state and is therefore always filled, whereas the level
$E_{+2}$ is always the highest energy state and is therefore always
empty. But the other two levels $E_{-2}$ and $E_{+1}$ could be filled
or empty depending on the relative values of $\Delta_Z$ and
$\Delta_{SAS}$, and in fact in the single particle model there is a
level crossing at $\Delta_Z=\Delta_{SAS}$ where these two levels
($E_{-2}$ and $E_{+1}$) are degenerate. Within the single particle
picture there will be a trivial first order phase transition in the
system at the $\Delta_Z=\Delta_{SAS}$ level crossing point ( as
$\Delta_Z$ and/or $\Delta_{SAS}$ are being varied as tuning
parameters) with the system making a transition from a fully
spin-polarized ferromagnetic (F) for $\Delta_Z>\Delta_{SAS}$, with the
occupation of both the spin-up $E_{-1}$ and $E_{-2}$ (symmetric and
antisymmetric) levels (with $E_+$ levels being completely empty) in
the high field (more precisely, the large Zeeman splitting) situation
to a fully pseudospin polarized and spin paramagnetic singlet state (S)
for $\Delta_{SAS}>\Delta_Z$, with the occupation of symmetric $E_{-1}$
and $E_{+1}$ spin up and down levels (with antisymmetric $E_{\pm2}$
levels being completely empty). There is nothing interesting or noteworthy 
about this trivial $\Delta_Z=\Delta_{SAS}$ level crossing induced first order 
phase transition except that {\it it does not happen} -- inclusion of many-body
interaction effects, particularly interlayer correlations, pre-empts the 
trivial first order transition by introducing a novel spin symmetry breaking 
\cite{citation5,citation6,citation7,citation8} which eliminates
the $\Delta_Z=\Delta_{SAS}$ induced level crossing by producing a new
symmetry-broken ground state which mixes the two relevant levels
$E_{+1}$ and $E_{-2}$ generating a new ground state with the linear
combination $\alpha |+1 \rangle + \beta |-2 \rangle$ (and $|\alpha|^2
+ |\beta|^2=1$).  The $\Delta_Z=\Delta_{SAS}$-induced level crossing
of spin-up/symmetric $|+1 \rangle$ and spin down/antisymmetric $|-2
\rangle$ states and the associated first order transition, therefore, does
not happen -- instead a new purely interaction driven quantum phase,
the so-called canted antiferromagnetic (C) phase \cite{citation6},
is stabilized around the $\Delta_Z=\Delta_{SAS}$ regime of the
relevant parameter space in between the F phase
($\Delta_{SAS}>>\Delta_Z$) and the S phase $\Delta_{SAS}>>\Delta_Z$.
This novel quantum phase diagram, where the phase transitions $F
\leftrightarrow C \leftrightarrow S$ are all continuous, was first 
predicted on the basis of an unrestricted Hartree-Fock mean-field
calculation in ref. \cite{citation5}, and was then further
theoretically extended and confirmed in refs 
\cite{citation6,citation7,citation8,citation9,citation10,citation11,citation12,citation13,citation14,citation15} using a variety of
theoretical techniques including the spin-bond approach
\cite{citation8,citation10,citation11,citation12,citation14}, more 
detailed Hartree-Fock calculations
\cite{citation7,citation9,citation10}, an $O(3)$ quantum non-linear
sigma model \cite{citation6,citation7}, a Chern-Simons field theory
\cite{citation13}, and a small system direct numerical diagonalization
calculation \cite{citation15}. It is reasonable to state that the
basic $\nu=2$ bilayer quantum Hall phase diagram with three distinct
spin quantum phases (F,C,S) as a function of interlayer tunneling,
Zeeman splitting, and interlayer correlation is now well established.
The spin orientations in the two layers in the C phase
are depicted schematically in Fig. 2.
\begin{figure*}[h]
\centerline{\epsfxsize=6cm
\epsfbox{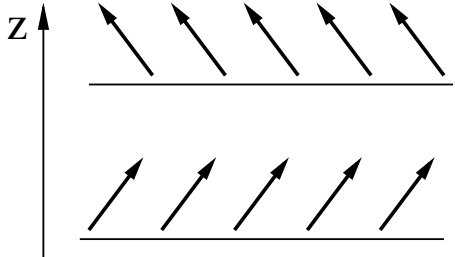}
}
\caption{Spin orientations in bilayer quantum Hall systems
at $\nu=2$ in the canted antiferromagnetic phase.}
\label{f2.eps}
\end{figure*}
Our description of the C phase and the associated $\nu=2$ bilayer quantum
Hall  phase diagram has so far utilized the single particle language
emphasizing the competition between $\Delta_Z$ and $\Delta_{SAS}$, and the
interaction-induced avoided level crossing at the 
$\Delta_Z = \Delta_{SAS}$
degeneracy point leading to the spontaneous breaking of bilayer spin
symmetry.  This is however purely qualitative and is in fact quite
simplistic since interaction strongly renormalizes the single particle
energy levels in the system.  One should interpret the energy levels of
Fig. 1 as renormalized (for example, at the Hartree-Fock level) levels, and
NOT purely single particle levels, i.e. the parameters $\Delta_Z$ and
$\Delta_{SAS}$ should not be taken as the bare parameters, but as the
renormalized effective spin and pseudospin splittings, respectively
(remembering that the many-body renormalization by Coulomb interaction may
be quite large in quantum Hall systems).  One implication of this many-body
renormalization is that the C phase may in fact extend all the way down to
$\Delta_Z \propto \Delta_{SAS}^2$ 
instead of being around $\Delta_Z \propto \Delta_{SAS}$ as the
simple level crossing qualitative argument above suggests.  Another feature
of the many-body nature of the phenomena manifests itself in the fact that
the quantum phase transitions among the F,C, and S phases may be induced by
tuning the interlayer separation d (instead of $\Delta_Z$ or $\Delta_{SAS}$ as
discussed above) which automatically continuously renormalizes the energy
levels producing the symmetry breaking.  Thus, $\Delta_Z$ and $\Delta_{SAS}$ 
in the
above discussion are not the bare single particle parameters, but are
renormalized effective parameters of the system, which could, in principle,
be very different from the bare parameters.

The spin symmetry breaking associated with the existence of the novel
C phase (not present in the noninteracting theory) leads to
interesting collective mode behavior
\cite{citation6,citation7,citation8} as one would expect for a
continuous quantum phase transition. First, the lower $S=1$ spin
density excitation, the $\omega_-$ mode associated with the collective
transition between $E_{-2}$ and $E_{+1}$, becomes soft with its long
wavelength spin gap vanishing at the boundaries between the C phase
and the F or S phase. This theoretically predicted long wavelength
softening of the appropriate spin density excitaion has been
experimentally observed \cite{citation16,citation17} via the inelastic
light scattering spectroscopy, confirming the existence of the C phase.
The second interesting collective mode behavior is the existence of
a Goldstone mode in the C phase with a linear energy-wavevector
dispersion in the long wavelength limit which arises from the
spontaneous breaking of the symmetry of spin rotations in the $XY$
plane in the CAF phase. This Goldstone mode, which exists only in the
CAF phase, has not yet been experimentally observed perhaps because of
the fact that the specific selection rules operating in the resonant
Raman scattering experiments produces little spectral weight in the
Goldstone mode making it unobservable in the usual inelastic light
scattering spectroscopy \cite{citation16,citation17}. The issue of the
observation of the Goldstone mode, which is a characteristic of the
CAF-phase, is an important open question in the subject.
It has been argued theoretically \cite{citation8} that disorder
may modify
the sharp Goldstone mode to a broad peak.

We close this introduction with a brief discussion of the nature of
the canted antiferromagnetic phase and the associated symmetry
breaking leading to it. In the absence of Zeeman splitting the spin
symmetry of the problem is $SU(2)$ whereas the pseudospin/layer index
has only a $U(1)$ symmetry in the absence of tunneling (the interlayer
interaction is explicitly different from the interlayer interaction,
the zero tunneling pseudospin symmetry becomes $SU(2)$ only in the
limit of zero interlayer separation when the the intralayer and
interlayer Coulomb interaction matrix elements are trivially equal). In
the presence of tunneling, however, the $U(1)$ pseudospin symmetry is
explicitly broken, and the only $SU(2)$ spin symmetry of the problem
matters. The presence of the external magnetic field along the $z$
direction in the quantum Hall problem now converts the whole problem
to that of an $XY$ quantum spin model. This $XY$ spin symmetry is
spontaneously broken in the C-phase as the electron spin chooses a
particular orientation in the $XY$ plane ( opposite or
antiferromagnetic in the two layers), making the total spin in each
layer to be canted (in opposite directions) at some angle to the $z$
axis (Fig. 2). The physical reason for this canting is simple: 
In the presence
of finite tunneling the spontaneous symmetry breaking induced
(antiferromagnetic) spin canting allows the $\nu=2$ bilayer system to
lower its total energy with respect to the normal F and S phases by
exploiting the superexchange or virtual exchange mechanism -- the
effect is akin to (but more complex than) the conversion of a Hubbard
type model to the $t$-$J$ model with $J=4t^2/U$, and the $J$ term
allows energy reduction via superexchange. The physical mechanism
underlying the spontaneous symmetry breaking leading to the C phase is
so transparent that one expects it to occur at the bilayer filling
factors $\nu=2/m$, where $m$ is an odd integer, provided there is
finite interlayer tunneling (and interlayer separation is neither too
large nor too small). The reason for this expectation is that for $m$
an odd integer, the $\nu=2/m$ bilayer system supports two
incompressible Laughlin states in each layer (when the layer
separation is large) which should exploit the superexchange mechanism
to lower its energy by creating the intermediate C-phase for
intermediate values of $\Delta_{SAS}$, $\Delta_Z$ and layer
separation. More detailed considerations based on the Chern-Simons theory
indicate that \cite{citation13}  C phases may exist even for general bilayer
filling factors not satisfying $\nu=2/m$ constraint.

The rest of this review is organized as follows: in section \ref{sec2} we briefly
describe the Hartree-Fock \cite{citation5,citation6,citation10} and the spin bond
mean field theories for the C-phase symmetry breaking; in section \ref{sec3} we
consider effects of disorder which, we argue theoretically \cite{citation8}, should
give rise to Bose spin glass phase in the quantum phase diagram -- there is a recent
experimental claim  of observing \cite{citation19} the predicted \cite{citation8}
Bose glass phase in $\nu=2$ bilayer transport experiments; in section
\ref{sec4} we discuss microscopic wavefunctions for the CAF phase
and the corresponding Chern-Simons field theory 
developed in ref \cite{citation13}; in section \ref{sec5} we conclude with a
discussion of open issues and questions and related theoretical developments
as well as a brief description of the experimental efforts in this problem
\cite{citation16,citation17,citation18,citation19,citation20}. 

\section{Mean Field Hartree-Fock and Spin Bond Theories}
\label{sec2}

The Hartree-Fock calculations for the CAF phase in $\nu=2$ bilayer
quantum Hall systems follow the usual approach of the mean field
theories.  One starts with the Coulomb interaction Hamiltonian
\begin{eqnarray}
{\cal H}_c &=& \frac{1}{2} \sum_{\sigma_1\sigma_2}
\sum_{\mu_1 \mu_2 \mu_3 \mu_4} \sum_{\alpha_1\alpha_2} 
%\frac{1}{\Omega}
\sum_{{\bf q}} V_{\mu_1\mu_2\mu_3\mu_4}({\bf q})e^{-q^2 l^2_0/2}
\nonumber\\
&\times&
e^{i q_x ( \alpha_1 - \alpha_2)l_0^2}
C^{\dagger}_{\alpha_1+q_y\mu_1\sigma_1}
C^{\dagger}_{\alpha_2\mu_2\sigma_2}
C_{\alpha_2+q_y\mu_3\sigma_2}
C_{\alpha_1\mu_4\sigma_1}
\label{H}
\end{eqnarray}
where $C^{\dagger}_{\alpha\mu\sigma}$ creates an electron in the
lowest Landau level with the intra-Landau index $\alpha$, pseudospin
$\mu$, and spin $\sigma$; and $ V_{\mu_1\mu_2\mu_3\mu_4}$ are 
the intralayer and interlayer Coulomb interaction
potentials (see \cite{citation5,citation6,citation7,citation9} for details).  
Mean field approximation to a
many-body problem is achieved by performing Hartree-Fock
pairing through the expectation values $\langle
C^{\dagger}_{\mu_1\sigma_1}C_{\mu_2\sigma_2} \rangle$, which should be
found self-consistently. It is natural to assume expectation values of
the fields that come into the non-interacting part of the Hamiltonian
$\langle C^{\dagger}_{\mu\sigma} C_{\mu\sigma} \rangle$.  Such terms
generate effective tunneling and Zeeman splitting that strongly
renormalize the bare parameters present in the non-interacting part of
the Hamiltonian. They provide the Hund's rule tendency of 
quantum Hall systems: by generating large effective Zeeman or
tunneling splitting all the electrons will be polarized in the spin or
pseudospin sectors in phases F and S respectively.  What is more
surprising is that one also finds a non-trivial self-consistent
solution for the fields $\langle C^{\dagger}_{1+}
C_{2-} \rangle$ and its conjugate. This field breaks
explicitly the $S^z$ spin symmetry of the original Hamiltonian
(\ref{H}) and provides mixing between the states $E_{-2}$ and
$E_{+1}$.  The existence of such non-trivial expectation values
substantiates the statment of the existence of spontaneously broken
spin symmetry in the CAF phase.  By going back to the basis of layer
indices for the single electron states (from the basis of
symmetric/antisymmetric states) one can show that $\langle
C^{\dagger}_{1+} C_{2-} \rangle$ describes the
appearance of the spin expectation values that lie in the $XY$ plane
and are opposite in the two layers, i.e. a canted spin phase.

An apparent disadvantage of the Hartree-Fock approach is that it does
not allow a consistent treatment of the quantum fluctuations of the
Neel order parameter, so it overestimates the region of stability of
the CAF phase and does not allow to address any of the interesting
critical phenomena that take place in the vicinity of the transition
lines.  Non-linear sigma model introduced in \cite{citation6,citation7} 
is a way to
address both of these issues in a simple field theoretical
approach. The two ingredients that come into this theory are: two
well separated layers form fully polarized ferromagnets with a gap
towards charge excitaions and the primary coupling between the layers
is antiferromagnetic exchange. The disadvantage of the non-linear
sigma model is that it treats the interlayer tunneling perturbatively
and by concentrating on the spin it does not provide an equivalent
treatment of the pseudospin degrees of freedom. So, for example,
understanding the effects of an external gate voltage would be extremely
difficult within this formalism. A spin bond approach suggested in \cite{citation8}
extends the non-linear sigma model by first treating
non-perturbatively the inter-layer Coulomb interactions, interlayer
tunneling, and the gate voltage to find the nature of the spin triplet and
spin singlet states that provide the basis for the effective
description of the system. In a nutshell the spin bond approach may be
summarized as follows. Two electrons with the same intra-Landau
level index but from the opposite layers may be combined into a spin
singlet or spin triplet electron pairs.  These combinations are then
treated as hard core bosons, with the interaction between the two
kinds of bosons coming from in-plane ferromagnetic exchange.  When
only the singlet or the triplet pairs are present in the ground state
one finds the S and the F phases respectively, and when the two bosons
are condensed simultaneously we find the CAF phase. There is a simple
connection between the Hartree-Fock and the spin bond formalisms.
In fact, if one considers a set of states that have the same expectation 
values as we discussed earlier, but allows these expectation values
to be non-uniform,
the energy functional for these states will coincide with the
continuous limit of the spin-bond model. 
Reasonable quantitative agreement between the spin-bond model
and the Hartree-Fock calculations has been demonstrated in 
\cite{citation8,citation12},
but computationally the first is significantly simpler.

\begin{figure*}[h]
\centerline{\epsfxsize=8cm
\epsfbox{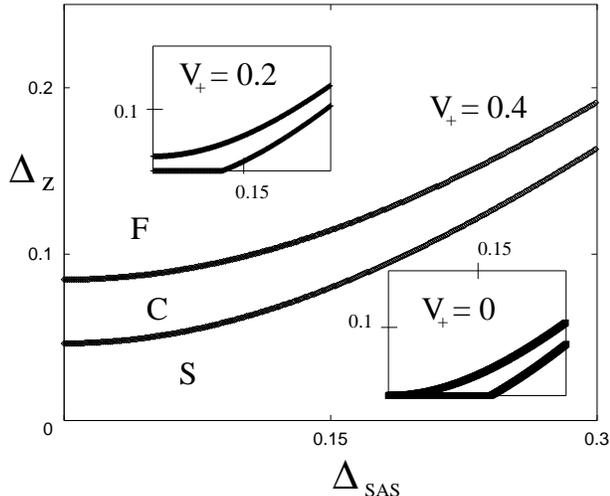}
}
\caption{Phase diagram for $\nu=2$ bilayer quantum Hall
system calculated using spin bond theory for different gate voltages. 
The length
and the energy units are the magnetic length $l_0$ and the interlayer
Coulomb energy $e^2/(\epsilon l_0)$. The interlayer separation is 1.
Phase diagrams for different gate voltage $V_+$ are shown.}
\label{fig2.eps}
\end{figure*}

The phase diagram obtained from the spin-bond model for the
case when the distance between the layers is equal to the magnetic
length $l_0$ is shown for three values of the external gate voltage
in Fig. 3. 
One can see that for large values of
$\Delta_{SAS}$ and $\Delta_{Z}$, the many body corrections to the
effective tunneling and Zeeman energies are small and we find that the
CAF phase is centered around $\Delta_{SAS}=\Delta_{Z}$ line as
discussed earlier. However when $\Delta_{SAS}$ and $\Delta_{Z}$ are
small, there are significant deviations from this line caused by
Coulomb interactions. Spin-bond model or Hartree-Fock approaches
allow one to calculate
several experimentally testable properties of the CAF phase.  Among
them is the amplitude of the antiferromagnetic order parameter $|{\bf
N}|$ that appears continuously on the phase boundaries between the C
and S or F phases and reaches its maximum in the midddle of the C
phase \cite{citation10}. From $|{\bf N}|$ one can calculate the Goldstone mode velocity
as well as find an estimate for the Kosterlitz-Thouless transition
temperature.  Another important probe of the CAF phase that has been
discussed in refs \cite{citation8,citation9} 
comes from applying a bias voltage ($V_+$) between
the two layers.  As shown in Fig. 3 the main effect of the charge
imbalance between the layers is to shift the phase boundaries without
altering their shape, so for example when $V_{+}$ is finite one can
find C and S phase even when $\Delta_{SAS}$ is zero.  In the context
of bilayer $\nu=1$ quantum Hall states a considerable emphasis has
been given to the concept of the interlayer coherent states \cite{citation22}. 
We 
emphasize that the concept of spontaneous interlayer coherence
is only applicable when there is no tunneling between the layers. So,
at $\nu=2$ only when $\Delta_{SAS}$ is identically zero and there is
finite bias voltage one can identify the CAF phase as
interlayer coherent, qualitatively similar to the pseudospin coherent
$(1,1,1)$ Halperin state for the $\nu=1$ bilayer system 
\cite{citation9,citation10}.

\section{Disorder Effects}
\label{sec3}

Another obvious advantage of the spin bond model is that
it gives a simple framework to understand the effects of disorder
on the C phase \cite{citation8}. Fluctuations in the distance between the wells or 
the presence of impurities give rise to random fluctuations
in the energy of singlet and triplet bosons, and will stabilize
the phase that may be visualized as consisting of domains of S, F, and CAF
phases. By appealling to a similarity between this problem
and a problem of charged bosons hopping in a random chemical potential
this phase was identified as a spin Bose glass phase \cite{citation8}.  
From the same analogy important conclusions may be drawn about the
properties of this novel phase. Triplet and singlet bosons are
localized, so there is no broken spin symmetry, however, there is
infinite antiferromagnetic susceptibility, analogous to infinite
superfluid susceptibilty of the charge Bose glass. Goldstone peaks, that
were the dominant feature of the spin response function in the CAF
phase, are replaced by finite longitudinal susceptibility at small
frequencies. This spin Bose glass phase has a finite density of low
energy excitations which provides another way to experimentally
distinguish it. Finally the existence of the spin Bose glass phase
separating the S, F, and CAF phases has important consequences in that
it changes the critical exponents to those of the
superconductor-insulator transition in the dirty boson system studied
in \cite{citation6,citation7}. It is worth pointing out that the spin Bose glass
system may be a better experimental realization of a 2D
superconductor-insulator transition in a boson system than
the conventionally used 2D superconducting films in that it is free of
long range forces and low energy fermionic excitations, and 
therefore allows one to vary
the density of bosons by varying the magnetic field.

\section{Halperin Wavefunction and the Chern-Simons Field Theory}
\label{sec4}

Another important perspective on the nature of 
the  C phase comes from considering its microscopic (Halperin)
wavefunction. Here again analogy with the spinless
electron in a bilayer system at $\nu=1$ (or $\nu=1/m$
in general) is very useful. In the latter case  Wen and Zee
\cite{Wen} discussed that the origin of the interlayer coherent
wavefunction at this filling factor lies in the
remarkable property of the Halperin $(m,m,m)$ wavefunctions
(neglecting electron spin) to fix the total filling
factor but not the individual filling factor
in each layer. This allows a construction of interlayer coherent states,
i.e. the states that
are a superposition of states with different numbers of particles
in each layer but fixed total number of particles. Such
states break spontaneously the $U(1)$ symmetry of the problem that in the 
absence
of interlayer tunneling the  number of electrons in each layer is
a good quantum number, and that the system should have a specific
number of electrons in each layer. For spinful electrons
in bilayer quantum Hall systems the analogous Halperin-type
construction leads naturally to a C phase \cite{citation13}. In fact imagine making
an analogue of the $(m,m,m)$ wavefunction of the $E_{-2}$ and $E_{+1}$
states. In this case one is mixing states that have different $S^z$ components
of the spin and therefore creates a state that is not 
an irreducible  represenatation
of the $S^z$ operator. This corresponds to spontaneously 
breaking the $S^z$ symmetry
of the system. From this reasoning one can also argue that the general
fractional filling factor for which the CAF phase is possible is given by
\begin{eqnarray}
\nu=\frac{n+m-2l}{nm-l^2}
\end{eqnarray}
where $n$ and $m$ are odd integers and $l$ is an arbitrary integer. This formula
trivially includes the $\nu=2/m$ originally suggested in 
\cite{citation6,citation7}.
The idea of the Halperin wavefunctions for 
the CAF phase can be extended to a full bosonic
Chern-Simons theory that provides a unified picture of the gapless
charge neutral Goldstone mode and charged excitations in the system 
\cite{citation13}.
The key ingredient of this approach is to represent the electrons
as bosons with  attached fluxe tubes and then describe the
quantum Hall state as a state where the bosons have condensed.
The main result that comes out of such calculation is 
that in the CAF phase the system
has non-trivial  topological vortex-like spin excitations, merons,
that have a Neel order parameter  winding by $2\pi$ on the periphery
with  S or F phases inside the meron core, and that these excitations carry
an electric charge. The amount of charge carried by each  meron depends on the exact
position inside the quantum phase diagram, but the two merons with  S and F
cores always add up to a charge of one quasiparticle 
(i.e. $1$ for the $\nu=2$ case). 
This provides another demonstartion of the remarkable property of quantum Hall
systems to mix charge and spin degrees of freedom \cite{citation4}.
Full experimental implications of the theory developed in \cite{citation13}
have not yet been worked out.

\section{Conclusions}
\label{sec5}

We have provided a brief qualitative review of recent developments in our
understanding of the rich quantum phase diagram of bilayer quantum
Hall systems at the total filling factor of $\nu=2$ or more generally
$\nu=2/m$ where $m$ is an odd integer.
Similar considerations should also apply to bilayer systems
with $\nu=2m$, where $m$ is odd, but Landau level coupling neglected here
may be a significant issue for $\nu=6$, $10$.
Interplay among interlayer
tunneling, Zeeman spin splitting, intra- and inter-layer Coulomb
interactions could produce novel spin symmetry breaking in the system,
leading to a new class of magnetic ground states, the canted
antiferromagnetic phase, nestled between the more usual spin-polarized
ferromagnetic and symmetric (singlet) paramagnetic phases. The
presence of disorder leads to interesting Bose glass regimes within
the canted phase. The spontaneous spin $XY$ symmetry breaking giving
rise to the canted phase also produces softening of
the appropriate collective spin density excitations in the F and S
phases, which have presumably been experimentally observed
\cite{citation16,citation17} via the resonant Raman scattering
spectroscopy. There have also been several reports of indirect
observation of the canted and the Bose glass phase in transport
experiments \cite{citation18,citation19,citation20}. There are
collective linearly dispersing Goldstone modes in the CAF phase which
have not yet been experimentally observed -- it has been argued
\cite{citation8} that the spectral weight carried by the Goldstone
mode will be broadened into a weak broad peak in the Bose glass phase,
which in fact is consistent with experimental light scattering
measurements Further work along this line is necessary.

In addition to the $T=0$ quantum phase transitions among the F, C, 
and S magnetic
phases, there should be a finite temperature classical 
Kosterlitz-Thouless type phase transitions 
\cite{citation5,citation6,citation7,citation14} within the CAF phase
as the temperature is increased, and the usual vortex unbinding
disordering transition of the $X-Y$ model takes place. The critical
temperature for this Kosterlitz-Thouless type transition
has been estimated \cite{citation7} to be around 1K, and there is some experimental
evidence in its support \cite{citation16,citation17}.

From a theoretical perspective the existence of the predicted canted phase
and the associated continuous quantum phase transitions in bilayer
quantum Hall systems is now well established. The original
prediction  \cite{citation5} based on an unrestricted Hartree-Fock mean field
theory has been well-confirmed and substantiated in subsequent theoretical
analysis using O(3) nonlinear sigma model  \cite{citation6,citation7},
spin bond theories  \cite{citation7,citation11,citation12},
more detailed Hartree-Fock calculations \cite{citation9,citation10}, and
most importantly, through the construction of explicit wavefunctions for
the symmetry-broken phases associated with a Chern-Simons theory \cite{citation13}
and a direct diagonalization exact numerical calculations \cite{citation15}.
The direct diagonalization calculations \cite{citation15} eliminates any lingering
questions one may have about the existence of the C phase being an artifact
of mean-field theories. The quantum phase diagram obtained in the exact
diagonalization calculation in ref. \cite{citation15} is 
essentially identical to that obtained \cite{citation8} by the spin bond
approach with the original Hartree-Fock theory 
\cite{citation5,citation6,citation7,citation10} giving phase diagrams
which are qualitatively identical except for the fact that the S phase
is overemphasized in Hartree-Fock theories.

The experimental situation is somewhat less definitive although the 
inelastic light scattering experiments \cite{citation16,citation17}
convincingly demonstrate the mode softening and the quantum phase
transition (and possibly even the classical Kosterlitz-Thouless transition) predicted
\cite{citation5,citation6,citation7} to occur in $\nu=2$ bilayer
quantum Hall systems. In addition, transport measurements of quantum Hall
activation energies provide strong circumstantial evidence \cite{citation18,citation19}
for the existence of the canted phase. More detiled transport measurements
on samples spanning wider regimes of parameter space (i.e. $\Delta_{SAS}$, $\Delta_Z$,
$d$, etc.) would certainly be helpful in elucidating the complete quantum 
phase diagram. In principle, definitive experimental evidence confirming
the canted phase could come from careful spin polarization and NMR measurements
as suggested in ref. \cite{citation10}, but such magnetic moment measurements
are extremely difficult in two dimensional semiconductor systems because of weak signal
and background problems. 

Finally we mention that there have been a number of related theoretical
developments  \cite{citation22,citation23,citation24,citation25,citation26}
in the field motivated by the prediction of the canted phase
\cite{citation5,citation6,citation7}. 
These theoretical developments explore several interesting
features such as generalizing the C phase concept to 
multilayer superlattice \cite{citation22} and to double quantum dot \cite{citation26}
structures as well as other topics in collective modes \cite{citation24},
the Kosterlitz-Thouless transition temperature \cite{citation23}, and the field
theory \cite{citation25} of the problem. A very recent preprint \cite{Fradkin}
considers the effects of spontaneous symmetry breaking in bilayer quantum Hall
systems on the edge states.

We also point out that mean field calculations \cite{DS} suggest the
possibility of an exchange-correlation-driven
intersubband-spin-density-softening-induced ground state antiferromagnetic
instability even for zero magnetic filelds in low density two-subband
(bilayer or monolayer) two-dimensional electron systems, however, Raman
scattering measurements indicate \cite{Plaut} no such zero field 
transitions for currently
accessible 2D densities.

We acknowledge US-ONR (S.D.S) and Harvard Society of Fellows (E.D) for support.

\end{document}